\documentclass{iaus}
\usepackage{graphicx}

\title[Preliminary results on SiO $v$=3 $J$=1--0 maser emission from AGB stars] 
{Preliminary results on SiO $v$=3 $J$=1--0 maser emission from AGB stars}

\author[J.-F. Desmurs, et al. ]   %% give here short author list %%
{J.-F. Desmurs$^1$, V. Bujarrabal$^1$, M. Lindqvist$^2$, J. Alcolea$^1$, R. Soria-Ruiz$^1$, \and P. Bergman$^2$}
\affiliation{$^1$Observatorio Astron\'omico Nacional, Madrid, Spain.
\\ email: {\tt [desmurs, bujarrabal, alcolea, r.soria] @oan.es } \\[\affilskip] 
$^2$Onsala Space Observatory, Chalmers Univ. of Technology, Sweden
\\email: {\tt [michael.lindqvist, pbergman] @chalmers.se}}

\pubyear{2012}
\volume{xxx}  %% insert here IAU Symposium No.
\pagerange{001--002}
\jname{Cosmic Masers - from OH to H{$_0$}}
\editors{Roy Booth, Liz Humphries \& Wouter Vlemmings Editor, eds.}
\begin{document}

\maketitle

\begin{abstract}
We present the results of SiO maser observations at 43~GHz toward two
AGB stars using the VLBA.  Our preliminary results on the relative
positions of the different $J$=1--0 SiO masers ($v$=1,2 and 3) indicate
that the current ideas on SiO maser pumping could be wrong at some
fundamental level. A deep revision of the SiO pumping models could be
necessary.  
\end{abstract}

\firstsection % if your document starts with a section,

\section{Introduction}

Many stars have been mapped in SiO emission $J$=1--0 $v$=1 and 2,
particularly using the VLBA (\cite[Diamond et al.\ 1994]{Diamond94},
\cite[Desmurs et al.\ 2000]{Desmurs00}, \cite[Cotton et al.\
2006]{Cotton06}, etc). The maser emission is found to form a ring of
spots at a few stellar radii from the center of the star. In general, both
distributions are similar, although the spots are very rarely
coincident and the $v$=2 ring is slightly closer to the star (see e.g.\
\cite[Desmurs et al.\ 2000]{Desmurs00}).

The similar distributions of the $v$=1, 2 $J$=1--0 transitions were
first interpreted as favoring collisional pumping, because the
radiative mechanisms tend to discriminate somewhat more strongly both
states. But the lack of coincidence was used as an argument in favor of
radiative pumping, leading to the well-known, long-lasting discrepancy
in the interpretation of the $v$=1, 2 $J$=1--0 maps in terms of pumping
mechanisms (see discussion in e.g.\ \cite[Desmurs et al.\
2000]{Desmurs00}).

The discussion on this topic has dramatically changed when the first
comparisons between the $v$=1 $J$=1--0 and $J$=2--1 maser distributions
were performed (see \cite[Soria-Ruiz et al.\ 2004, 2005, 2007]{Soria04,
Soria05, Soria07}). In contradiction with predictions, from both
radiative and collisional models, the $v$=1 $J$=2--1 maser spots
systematically occupy a ring with a significantly larger radius
($\approx$30\%) than that of $v$=1 $J$=1--0, both spot distributions
being completely unrelated. Soria-Ruiz et al.\ (2004) interpreted these
unexpected results invoking line overlap between the ro-vibrational
transitions $v$=1\,$J$=0 -- $v$=2\,$J$=1 of SiO and $\nu_2$=0\, $J_{k_a
  k_c}$=12 $_{7,5}$ -- $\nu_2$=1\, $J_{k_a k_c}$=11$_{6,6}$ of H$_2$O.
%%vb
This phenomenon would also introduce a strong coupling of the $v$=1 and
$v$=2 $J$=1--0 line, explaining their similar distribution.

If our present theoretical ideas are correct (e.g. Bujarrabal \&
Nguyen-Q-Rieu 1981, Bujarrabal 1994, Locket \& Elitzur 1992, Humphreys
et al. 2002), the $v$=3 $J$=1--0 emission should require completely
different excitation conditions than the other less excited lines. No
pair of overlapping lines is expected to couple the $v$=3 $J$=1--0
inversion with any of the other SiO lines. The $v$=3 $J$=1--0 spatial
distribution should be different compared with to the $J$=1--0 $v$=1, 2 ones
and, of course, of the $J$=2--1 $v$=1 maser, and placed in a still inner
ring than $v$=2.

\section{Preliminary results toward R Leo and TX Cam}

\begin{figure}[h]
\begin{center}
 \includegraphics[width=2.63in]{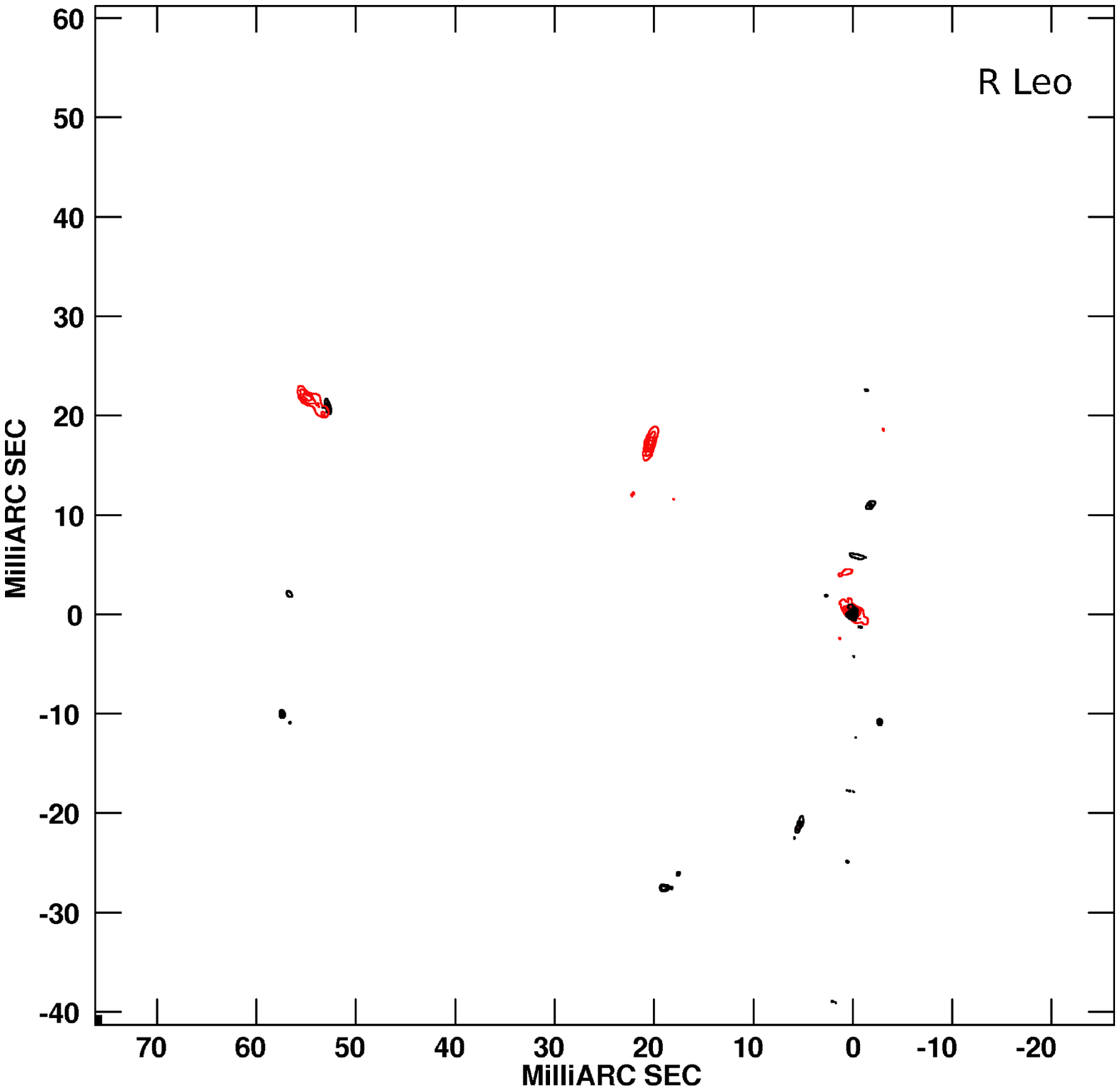}
 \includegraphics[width=2.63in]{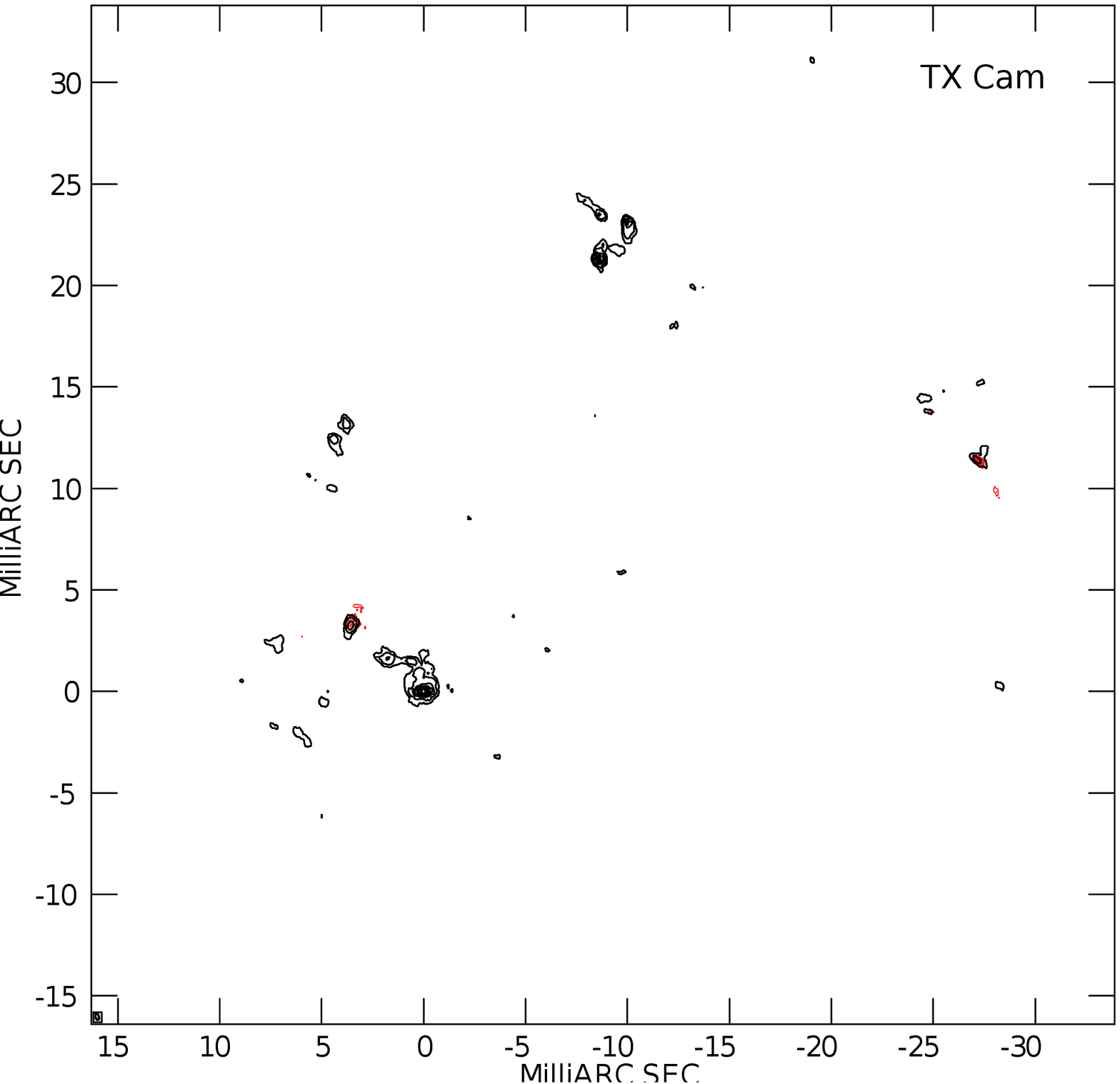}
 \caption{VLBA map of SiO $J$=1--0 $v$=2 (in black) and $v$=3 (in red)
 maser emission from R Leo (right) and TX Cam (left)}
   \label{fig1}
\end{center}
\end{figure}

The $v$=3 $J$=1--0 line is sometimes quite intense (\cite[Alcolea et
al.\ 1989]{Alcolea89}), and bright enough to be mapped with the VLBA,
but it is strongly variable, both in time (with characteristic times
scales of a few months) and from object to object.  With the 20-m
antenna of Onsala, we monitored a list of AGB stars to select the best
candidates to be mapped with the VLBA, observing simultaneously the
$v$=1,2 and 3 $J$=1--0 SiO masers at 42-43 GHz.

In the figure above, we show a preliminary map of the brightness
distribution of $^{28}$SiO $v$=2 and 3, $J$=1--0 obtained toward R Leo
(on the left) and TX Cam (on the right). These are the first VLBA maps
ever of the $v$=3 $J$=1--0 maser (in red). Although the alignment
between the maps of the two lines is just indicative (the observations
were not done in phase referencing mode, and the proposed alignment is
based on the similitude in velocity and the spatial distribution of
some spots). 

These preliminary results show a surprising similar distribution in
the $v$=3 $J$=1--0 and $v$=1, 2 $J$=1--0 masers. Would this result be
confirmed, our ideas on SiO maser pumping scheme must be wrong at some
fundamental level and a deep revision of the SiO pumping models will be
necessary.

%\begin{discussion}
%
%\discuss{}{}
%
%\end{discussion}

\end{document}